\documentclass[preprint]{aastex}

\newcommand{\fl}{_{\mbox{\tiny flat}}}
\newcommand{\bg}{_{\mbox{\tiny o}}}
\newcommand{\FF}{_{\mbox{\tiny FF}}}
\newcommand{\tl}{_{\mbox{\tiny tot}}}
\begin{document}


\title{An empirical model for protostellar collapse}


\author{A. P. Whitworth and D. Ward-Thompson}
\affil{Department of Physics \& Astronomy, Cardiff University, PO Box 913,
Cardiff, UK}

\begin{abstract}
We propose a new analytic model for the initial conditions of protostellar
collapse in relatively isolated regions of
star formation. The model is non-magnetic, and
is based on a Plummer-like 
radial density profile as its initial condition. It
fits: the observed density profiles of 
pre-stellar cores and Class 0 protostars; recent observations 
in pre-stellar cores of roughly constant contraction velocities
over a wide range of radii; and the lifetimes and accretion 
rates derived for Class 0 and Class I protostars. However, the model is
very simple, having in effect
only 2 free parameters, and so should provide a useful 
framework for interpreting observations of pre-stellar cores and protostars,
and for calculations of radiation transport and time-dependent chemistry.
As an example, we model the pre-stellar core L1544.
\end{abstract}


\section{Introduction}

Observations of the earliest stages of pre-stellar and protostellar 
evolution constrain the parameter space within which realistic models 
of low-mass (M$\stackrel{<}{\sim}$2--3M$_\odot$)
star formation can operate. The pioneering work of Myers and collaborators 
(e.g. Myers \& Benson 1983; Benson \& Myers 1989) and 
subsequent searches (e.g. Clemens \& Barvainis 1988; Bourke, Hyland 
\& Robinson 1995; Jessop \& Ward-Thompson 2000) have identified 
many dense, isolated cores in which star formation appears to have recently
taken place or to be about to occur. These cores can be divided into those 
with protostars, as deduced from their IRAS emission, and those without -- the 
`starless cores' (Beichman et al 1986). 
Subsequent study of submillimetre continuum emission from the cold dust 
in starless cores has identified those which are closest to forming 
protostars, which are labelled pre-stellar cores (Ward-Thompson et al 
1994; 1999). Pre-stellar cores are believed to represent imminent or ongoing
collapse, before a central star-like object has formed. Therefore 
they should reflect most closely the initial conditions for collapse.

Once a central star-like object forms, the source is known as a protostar. 
It is classified 
as a Class 0 protostar as long as most of the mass is still in the infalling 
envelope -- rather than in the central star-like object plus its accretion 
disc (Andr\'e, Ward-Thompson \& Barsony 1993 -- hereafter AWB93). 
After half the mass of the envelope has fallen in, 
the source is referred to as a Class I protostar
(Lada \& Wilking 1984; Lada 1987). Eventually most of the 
envelope has been accreted, or dissipated, and 
the source is referred to as a Class II source, which is a 
classical T Tauri star with a 
circumstellar disc. In the final pre-main-sequence stage 
the inner part of the disc has disappeared, and
it becomes a Class III source, which is a weak-line T Tauri star
(Andr\'e \& Montmerle 1994).
Any satisfactory model of star formation must reproduce the observed 
properties of these protostars, starting from the initial conditions 
observed in pre-stellar cores. In particular, we note the following 
apparently generic features:

\begin{enumerate}

\item Density profiles: Pre-stellar cores have flat central density 
profiles ranging from constant density to $\rho \propto r^{-1}$ 
in their innermost few thousand AU
(Ward-Thompson et al. 1994; Andr\'e, Ward-Thompson \& Motte 1996;
Ward-Thompson, Motte \& Andr\'e 1999). The density profiles then 
steepen towards their edges, demonstrating gradients up to $\rho \propto 
r^{-4}$ or even $r^{-5}$ at radii $\stackrel{>}{\sim}$10,000~AU
(Abergel et al. 1996; Bacmann et al. 2000). A single-index power-law
may fit a portion of the density profile of a given core, but no core
can be fitted by a single power-law throughout.

\item Time-scales:
Proceeding chronologically, the pre-stellar core phase 
lasts a few 10$^6$ years. The Class 0 protostellar 
phase then lasts a few 10$^4$ years, and is 
characterized by a very high accretion rate $\,(\stackrel{>}{\sim} 
10^{-5} M_\odot$ yr$^{-1})\,$, which delivers at least half of the 
final stellar mass. Finally, the Class I protostellar phase lasts 
a few 10$^5$ years, and is characterized 
by a significantly lower accretion rate $\,(\stackrel{<}{\sim} 10^{-6} 
M_\odot$ yr$^{-1})\,$, which delivers the remainder of the final stellar 
mass. We note that there is a large uncertainty 
in these time-scales, because they are derived from statistical 
arguments (e.g. Beichman et al. 1986; Andr\'e 1994; Ward-Thompson 1996;
Andr\'e, Ward-Thompson \& Barsony 2000 -- hereafter AWB00) 
and the statistics may be influenced by selection effects.

\item Velocity fields: There are already significant inward velocities 
in the outer layers of the pre-stellar core L1544, even though there is no 
evidence for a central protostar having formed yet (Tafalla et al. 1998;
Williams et al. 1999). Inward velocities have now
also been detected in several other pre-stellar cores
(e.g. Lee, Myers \& Tafalla 1999; Gregersen \& Evans 2000), and 
it appears that a few per~cent of cores may already be contracting.

\end{enumerate}

These observations cast considerable doubt on the star formation model 
proposed by Shu, Adams \& Lizano (1987) in which a singular isothermal 
sphere (SIS) 
undergoes inside-out collapse and delivers a constant accretion 
rate of order $a^3/G$ onto the central protostar (where $a$ is the isothermal 
sound speed). Additionally, there are many strong theoretical reasons 
(e.g. Whitworth et al. 1996) for questioning the relevance of the SIS model. 
In  particular, there is a low likelihood that nature can assemble anything 
approaching a SIS in the first place, due to its 
innate instability (e.g. Whitworth \& Summers 1985). Also, the 
pre-disposition of a SIS to collapse to a single star 
contrasts starkly with the large fraction of newly-formed 
stars observed 
to be in binary systems (e.g. Simon et al. 1995). 
Therefore it is appropriate to look for alternative models.
In this paper we present a simple analytic model for protostar formation 
satisfying the above observations. 
Our model is non-magnetic, and therefore provides a useful complement 
to magnetic models, such as those of
Safier, McKee \& Stahler (1997) and Ciolek \& Basu (2000).

However, not all of the observations present a coherent picture.
There is a well-known `luminosity problem' for 
Class 0 \& I sources (e.g. Kenyon et al. 1990),
namely that their low luminosities imply 
low accretion rates if the material is accreting directly onto the 
protostellar surface. Low accretion rates imply long lifetimes, which 
appear to be incompatible with the relatively small number of known Class 
0 \& I sources. One possible explanation in the case of Class I sources
is that the accretion rate remains low throughout 
the Class I phase, and most of the remaining protostellar 
envelope is dissipated in some way other than by being accreted 
onto the central protostar. For instance, it might be converted 
into lower-mass companions, as in the simulations of Boffin et al.
(1998) and Watkins et al. (1998a \& b). 

Alternatively, Kenyon \& Hartmann (1995) explain it by arguing that 
accretion takes place in a fairly uniform manner from the envelope onto
a disc, and is then episodic in nature
from the disc onto the central protostar. Short-lived (and therefore 
infrequently observed) rapid accretion episodes (manifesting themselves
as FU Orionis activity) are interspersed with long periods of slow, 
or even negligible, accretion during which matter piles up 
at the inner edge of the accretion disc. Thus protostars spend the
majority of their time in a `lower luminosity' phase, only flaring
up briefly when the accretion disc becomes unstable and matter is
rapidly deposited onto the protostar.

This explanation was used by Kenyon \& Hartmann (1995)
to account for the low luminosity of Taurus
Class I sources, but it could equally well apply to the Class 0 phase.
In our model we treat the central `protostar+disc' system simply as a
point source, since in most relevant mm/submm observations the protostar
and surrounding disc are unresolved, and we simply consider accretion
from the envelope onto the central `protostar+disc' system.
Episodic accretion from disc onto protostar solves the `luminosity
problem' for our model as happily as for any other star formation model.
In a future version of the model
we plan to include the effects of angular momentum, at which point we will
be able to model the disc separately from the central protostar (c.f.
Cassen \& Moosman 1981; Terebey, Shu \& Cassen 1984).

The paper is laid out as follows. Section 2.1 describes the mathematical
model. Section 2.2 details how the above observations constrain the model
parameters. Section 3 gives a specific example of using the model to
compare with one particular source, L1544: Section 3.1 lists the relevant
observations of L1544; Section 3.2 illustrates the specific model; Section 
3.3 compares the predictions of this model with measured lifetimes of other
pre-stellar cores and protostars. Section 4 compares our model to previously
published star formation models. Section 5 presents our main conclusions.

\section{The model}

\subsection{General analysis}
 
The model assumes that when a pre-stellar core becomes unstable 
against collapse at time $t = 0\,$, it is static and approximates to 
a Plummer-like density ($\rho$) profile (Plummer 1911), in terms of radius 
$r$, of the form

\begin{equation}
\label{RHOINIT}
\rho(r,t=0) \; = \; \rho\fl \left[ \frac{R\fl}
{\left( R\fl^2 + r^2 \right)^{1/2}} \right]^\eta .
\end{equation}

\noindent
Thus the initial density is approximately uniform at $\sim \rho\fl$ 
for $r \ll R\fl$, and it falls off as $r^{-\eta}$ for $r \gg R\fl$. 
We note parenthetically that the true Plummer Sphere, which 
is widely used in work on star and galaxy clusters, corresponds 
to the specific case $\eta = 5$. It is for this reason that we 
describe equation 1 as `Plummer-like'.
We choose this form, not because it is predicted by detailed 
thermodynamic considerations or simulations of core formation,
but because it seems to capture -- with a minimum number 
of free parameters -- the essential {\em observed} properties of 
pre-stellar cores. It is mathematically simple and -- as we now show --
it leads to predictions which accord well with the observational constraints.

We compute the subsequent evolution of this model pre-stellar core
on the assumption that it has negligible pressure, and therefore undergoes
freefall collapse, starting at time $t = 0$. For this purpose,
it is convenient to adopt a Lagrangian formulation, i.e. to switch 
from $(r,t)$ to $(M,t)$ as the independent variables. Given the initial 
density profile $\rho(r,t=0)\;$ in equation (\ref{RHOINIT}), we can integrate 
to obtain the initial mass profile

\begin{equation}
M(r,t=0) \; = \; \int_{r'=0}^{r'=r} \rho(r',t=0) \, 4 \pi r'^2 dr' \; .
\end{equation}

\noindent
This can then  be inverted to obtain $r(M,t=0)$, i.e. the initial 
radius of the shell with mass $M$ interior to it. The freefall time 
of the shell is

\begin{equation}
t\FF(M) \; = \; \frac{\pi}{2} \; \left[ \frac{r^3(M,0)}
{2 G M} \right]^{1/2} \; ,
\end{equation}

\noindent
and at this time the shell delivers a rate of infall

\begin{eqnarray} \nonumber
\dot{M}(M) & = & \left[ \frac{dt\FF}{dM}(M) \right]^{-1} \;  \\ \nonumber
 & & \\ 
 & = & \frac{64 G M^2 \rho(M,0) t\FF(M)}
{\pi \left[ 3 M - 4 \pi r^3(M,0) \rho(M,0) \right]} \;  \\ \nonumber
\end{eqnarray}

\noindent
onto the centre.
To obtain the density profile at an arbitrary (positive) time $t$, we 
can use the equation for freefall from rest. For all shells having 
$t\FF(M) > t$, the instantaneous 
radius $r(M,t)$ is given by

\begin{equation}
\frac{t}{t\FF(M)} \; = \; \frac{2}{\pi} \, \left\{ 
\mbox{cos}^{-1} \left( \left[ \frac{r(M,t)}{r(M,0)} 
\right]^{1/2} \right) \; + \; 
\left[ \frac{r(M,t)}{r(M,0)} \right]^{1/2} 
\left[ 1 - \frac{r(M,t)}{r(M,0)} \right]^{1/2} \right\} \; .
\end{equation}

\noindent
Knowing $r(M,t)$, we can calculate the instantaneous density profile from

\begin{equation}
\rho(M,t) \; = \; \left\{ 4 \pi r^2(M,t) 
\left. \frac{\partial r}{\partial M} \right|_t \right\}^{-1} \; .
\end{equation}

\noindent
We can also compute the instantaneous velocity profile from

\[
v(M,t) \; = \; \left\{ \frac{2 \, G \, M \, 
\left[ r(M,0) - r(M,t) \right]}
{r(M,0) \, r(M,t)} \right\} \; ,
\]

\noindent
and the instantaneous column-density at impact parameter $b$ from

\[
\Sigma(b,t) \; = \; \frac{1}{2 \pi} \, 
\int_{M=M(b,t)}^{M=M_{\mbox{\tiny tot}}} \, 
\frac{dM}{r(M,t) \, \left[ r^2(M,t) - b^2 \right]^{1/2}} \; .
\]

\noindent
The simplest computational strategy is to generate a dense table of 
values of $r(m,0)$, $\rho(M,0)$, $M$, $t_{\mbox{\tiny FF}}(M)$, 
$\dot{M}(M)$; and a second dense table of values of $\xi \equiv 
r(M,t)/r(M,0)$ and $\tau \equiv t / t_{\mbox{\tiny FF}}(M)$; then 
to use these for interpolation.

\subsection {Observational constraints}

In order to reproduce the observations described above, we must adopt 
$\eta \stackrel{>}{\sim} 4$. The choice of $\eta$ is dictated by the 
relative durations of the 
Class 0 and Class I protostellar phases, and the fact that at least 
half the final stellar mass is delivered during the Class 0 phase
(see AWB00 and references therein).
If a much smaller value of $\eta$ were adopted, the accretion rate 
would not 
decrease sufficiently rapidly going from the Class 0 to the Class I 
protostellar phase, i.e. by about an order of magnitude between the 
infall of the first half of the envelope mass (the Class 0 phase) 
and the infall or dissipation of the rest (the Class I phase). 
Indeed, for $\eta \leq 3$, 
the mass of the protostellar core diverges, and the accretion 
rate falls off less rapidly than t$^{-1}$. 

If a much larger value of $\eta$ were adopted, the accretion rate would 
fall off very rapidly, but this would only happen after almost all 
the envelope mass had fallen in; in other words, the 
accretion rate would switch rather abruptly from being 
substantial to being negligible. This would only be acceptable if 
stars assembled most of their mass rapidly during the Class 0 
phase and right at the beginning of the Class I phase, with the rest 
of the Class I phase contributing very little to the final mass. 
This may not be completely ruled out by the observations, particularly 
if the outer envelopes of protostars are dispersed rather than 
being accreted, but it would be difficult to reconcile with the
luminosities of Class I protostars. Hence, we conclude that the optimum
value of $\eta$ is 4, and we fix it at this value; therefore it is no longer
a free parameter of the model.

The value of $\rho\fl$ (the central density at the onset of freefall 
collapse) is dictated not only by observations of pre-stellar cores, 
but also by the duration of the Class 0 phase of a few 10$^4$ years. 
With $\eta = 4$, the duration of the 
rapid accretion phase (i.e. the Class 0 phase) is comparable to the 
initial freefall time at the centre of the pre-stellar core, 
$t\bg \equiv t_{FF}(M=0)$, 
so we require

\begin{equation}
\label{FREEFALL-TIME}
t\bg \; = \; \left[ \frac{3 \pi}{32 G \rho\fl} \right]^{1/2} 
\; \sim \; {\rm few} \; 10^4 \mbox{ years} \; , 
\end{equation}

\noindent
which gives $\rho\fl \sim 3 \times 10^{-18} \, \rightarrow \, 
3 \times 10^{-17}$ g cm$^{-3}$.
Finally, $R\fl$ is fixed by the total mass of the pre-stellar core 
being modelled.

\section{Detailed results and comparison with L1544}

\subsection{L1544 observations}

For the purposes of illustration, we have chosen 
to model the evolution of the pre-stellar core L1544, because it is 
one of the most carefully observed -- and hence one of the 
most tightly constrained -- pre-stellar cores. Its velocity field has 
been studied in detail (Tafalla et al. 1998; Williams et al. 1999). 
Its density profile and total mass ($M\tl \sim 8 M_\odot$) 
have been deduced from millimetre continuum observations 
(Ward-Thompson et al. 1999). And its far-infrared emission 
(as observed by the Infrared Space Observatory, ISO) shows it 
to be roughly isothermal (AWB00). 
We note that since our model is driven purely by self-gravity, it 
is very simple to re-scale the results to fit other sources, by 
altering $\rho\fl$ and $R\fl$. The masses then scale as 
$M\propto\rho\fl R\fl^3$, 
time-scales as $t\propto\rho\fl^{-1/2}$, 
velocities as $v\propto\rho\fl^{1/2}R\fl$, 
and accretion rates as $\dot{M}\propto\rho\fl^{3/2}R\fl^3$.

Tafalla et al. (1998) infer that the central core in L1544 is pre-stellar 
but probably close to forming a Class 0 protostar. From a detailed 
analysis of the profiles of various molecular 
lines -- in particular CS $(J = 2 \rightarrow 1)$, which is self-absorbed, 
and C$^{34}$S $(J = 2 \rightarrow 1)$, which is not -- they argue that 
the outer layers of L1544 are falling inwards. Specifically, they deduce
that gas at radii $r \sim 1-4 \times 10^4$ AU
is moving inwards at speed $v \sim 0.08$ km s$^{-1}$. 
Williams et al. (1999) detect similar infall speeds in their 
interferometric observations of N$_2$H$^+$ (J = 1$\rightarrow$0),
and infer gas
infalling at $\sim 0.08$ km s$^{-1}$ on much smaller
scales of $\sim$ 2,000 AU. We shall refer to the region 2,000 AU 
$\stackrel{<}{\sim} r \stackrel{<}{\sim}$ 40,000 AU as `the inflow region'.

Based on these observations, Williams et al. (1999) claimed that L1544 is
inconsistent with ambipolar diffusion models. Ciolek \& Basu (2000) then 
presented a model that they claimed could explain the observations, by taking 
initial conditions that are close to magnetically critical. 
SCUBA polarimetry observations of L1544
(Ward-Thompson et al. 2000) showed a magnetic field that does not align with 
the short axis of the core. This can also
be explained by initial conditions close to magnetically critical.
However, Bacmann 
et al. (2000) presented ISOCAM data that show that L1544 has a 
steep radial density profile in its outer edges. 
This requires highly magnetically sub-critical
initial conditions to explain it. 
Clearly there is a contradiction here.
One possible explanation is that the magnetic field may not be
playing a significant role, and we are simply seeing gravitational
contraction prior to the formation of a protostar.
Thus we shall attempt to model the above mentioned data of L1544, 
without invoking a magnetic field,
using $\eta = 4$, $\rho\fl = 3 \times 
10^{-18}$ g cm$^{-3}$ and $R\fl = 5350$ AU ($\equiv 0.026$ pc), 
giving a total core mass of $M\tl = 8 M_\odot$.
We stress that these values are chosen to match this one particular core.
Our presumption is that other cores have a range of values of 
$\rho\fl$ and $R\fl$.

\subsection{Model predictions for L1544}

Figure 1 shows the evolution of the volume density profile, 
through the end of the pre-stellar core phase
and the start of the Class 0 phase, as a log-log plot of 
number density (in H$_2$ per cm$^3$) versus radius 
(in AU). The profiles at different epochs are labelled according 
to the amount of time which has elapsed as a fraction of the 
initial free-fall time at the centre of the pre-stellar core, 
$t\bg$. Hence $(t/t\bg) = 0$ represents the initial conditions; 
and $(t/t\bg) = 1$ represents the profile after one 
central free-fall time, at the birth of the central point-mass --
i.e. the end of the pre-stellar phase and the start of the Class 0 phase. 

The dashed line shows the measured radial density profile of L1544 deduced 
from millimetre continuum observations (Ward-Thompson et al. 1999). 
It can be seen from Figure 1 that the model density profile starts  
relatively flat in the centre, and steepens towards the edge, in 
approximate agreement with the observed radial density profile. 
The crosses mark the density and radius at the positions of the
detected inflow,
as deduced by Tafalla et al. (1998) and Williams et al. (1999). 
These are also consistent with the model predictions at early times.

As the model evolves, it can be seen that the central density slowly 
increases. As one free-fall time is approached the central density 
profile starts to steepen, asymptotically approaching 
$\rho \propto r^{-3/2}$.
The outermost profile stays virtually unchanged at its initial 
$\rho \propto r^{-4}$.
After the formation of a central point-mass (at one 
free-fall time), the magnitude of the density decreases 
slowly as more of the circumstellar 
material accretes onto the central point-mass, but 
the shape of the inner
density profile stays similar at $\rho \propto r^{-3/2}$.
This is consistent with the recent results of
Chandler \& Richer (2000), who conclude on the basis of SCUBA images
that young Class 0 protostars approximate to $\rho \propto r^{-3/2}$
on scales of 300--10,000AU.

Figure 2 shows the evolution of the inward radial velocity profile, 
through the pre-stellar stage and the start of the Class 0 phase,
as a log-log plot of velocity $-v$ (in km s$^{-1}$) versus radius 
(in AU). Once again, 
the profiles at different epochs are labelled according 
to the amount of time which has elapsed as a fraction of the 
initial free-fall time at the centre of the pre-stellar core, 
$t\bg$. The initial conditions are static, so the velocity is zero everywhere 
at t=0. Between times of roughly $(t/t\bg) = 0.3$ to 
$(t/t\bg) = 0.7$ (the middle of this range is shown), 
the central velocity profile approximates to 
$v \propto r$, and the outer 
velocity profile approximates to $v \propto r^{-2}$.
Throughout this phase the absolute 
value of velocity increases monotonically with time at all radii.

After the formation of a central protostar at one free-fall time the 
central velocity profile tends towards $v \propto r^{-1/2}$.
However, the outer velocity 
profile remains relatively unchanged in form, 
although it carries on increasing in magnitude. 
The crosses on Figure 2 represent the velocity and radius at the two
positions in 
the inflow region deduced by Tafalla et al. (1998) and Williams 
et al. (1999) from their respective observations. They are consistent 
with the model predictions at $(t/t\bg) \simeq 0.5$.

Figure 3 shows how the accretion rate varies with time measured from 
the onset of freefall collapse, now plotted on a linear scale of $\dot M$ 
versus $t$. We note that the shape of this curve is determined 
solely by the choice of $\eta$. The choices of $\rho\fl$ and $R\fl$  
only influence the scaling. For any flat-topped initial density 
profile, there is a pause between the onset of freefall collapse at $t=0$, 
and the onset of accretion onto a central point-mass at $t=t\bg$
(remember that we are treating the central protostar and its circumstellar
disc merely as a point source).
At this point
the contracting pre-stellar core phase ends, the accretion rate
becomes non-zero and the object becomes a Class 0 source.

Hence this detailed comparison, using the specific example of
L1544, shows remarkably good agreement between our model and all of the
measured quantities. The apparently roughly uniform velocity profile is
well reproduced by a collapsing Plummer sphere at $t/t\bg \simeq 0.5$.
The measured density profile, both from the continuum and the
spectroscopy, is also consistent with $t/t\bg \simeq 0.5$.
Thus the pre-stellar core L1544 is consistent with our model,
in the earliest stages of collapse.

\subsection{Comparison of lifetimes}

We can compare the various lifetimes predicted by the model illustrated
in Figure 3 with those derived observationally from statistical arguments
(although we note that such arguments carry large uncertainties).
We have defined the time $t=0$ to be the point at which a pre-stellar core
begins to contract. We will refer to the time from $t=0$ until $t=t\bg$ as
the `contracting pre-stellar core phase'. Note that
this is not the same as the total
pre-stellar core phase, because most pre-stellar cores
show no evidence of infall motions
and therefore are presumably not contracting.
Our model says nothing about the total pre-stellar core lifetime,
merely about the lifetime of the contracting pre-stellar core phase.
Ward-Thompson et al. (1994)
estimated the lifetime of pre-stellar cores detected in the submm continuum
to be $\sim$10$^6$ years. This was based on a total starless core
lifetime of a
few 10$^6$ years, derived from the survey of Beichman et al. (1986).

Lee, Myers \& Tafalla (1999) surveyed a total of 220 starless and
pre-stellar cores and found definite evidence for collapse in only 7 of 
those. There was tentative evidence for contraction in a further 10 cores,
but there was also some doubt about the exact frequencies of some of the
transitions used in that survey, calling into question some of the tentative
detections, so we do not consider them here. If the fraction of contracting
pre-stellar cores is 7 out of 220, as suggested by Lee et al. (1999),
then the contracting pre-stellar core phase lasts
for only 3\% of the total starless and
pre-stellar core phase. Hence the lifetime of
the contracting pre-stellar core phase is
$\sim$3--10 $\times$ 10$^4$ years.
It can be seen from Figure 3 that our model predicts a value for this
phase consistent with the lower end of this range.

Gregersen \& Evans (2000) observed a subset of sources from the list of
Ward-Thompson et al. (1994), which in turn was drawn from the catalogue
of starless cores listed by Beichman et al. (1986), and discovered
six more cores with line asymmetries indicative of collapse. The list of
Beichman et al. (1986) contained almost 50 starless cores, hence the
result of Gregersen \& Evans (2000) appears to be implying a lifetime for
the contracting pre-stellar core phase of $\sim$10\% of that of the
total starless core phase, or a few 10$^5$ years. This is longer than
our model predicts.

However, the fact that two different statistical studies can produce
results for this timescale that differ by an order of magnitude 
illustrates the uncertainties inherent in such calculations. For example,
the Gregersen \& Evans (2000) sample contains
a great deal of biassing and selection effects, since they
consciously chose sources which were already believed to be
closest to forming Class 0 protostars. Hence an order of magnitude 
uncertainty in a statistical lifetime derived from such a sample is not
surprising. Nonetheless, if a significantly larger number of contracting
pre-stellar cores is discovered with improving detection techniques, and
the total pre-stellar core life-time remains unchanged, then
our model would have to be modified to increase the contracting pre-stellar
core lifetime (see section 4 below).

After the pre-stellar core stage
a central point-mass forms, and the accretion rate increases 
very rapidly to its maximum value, before decreasing on a time scale 
$\sim t\bg$. We have marked on Figure 3 the points at which various 
fractions of $M\tl$ have been accreted. Remember that the Class 0/I
borderline occurs at $M/M\tl=50\%$. It can be seen that the Class 0
phase lasts for $\sim$8$\times$10$^4$ years, and
has much more rapid accretion than the Class I phase.

We can compare the Class 0 lifetime predicted by our model with the 
lifetimes of various sources that have been estimated from observations.
AWB93 defined the Class 0/I borderline as the point where a source has
accreted half of its final main-sequence mass. They used the ratio between 
the submillimetre luminosity L$_{submm}$ and the bolometric luminosity 
L$_{bol}$ as an observational indicator of this, where Class 0
sources have L$_{submm}$/L$_{bol}>$ 0.005, and Class I sources have
L$_{submm}$/L$_{bol}<$ 0.005.

VLA1623 has L$_{submm}$/L$_{bol}\sim$0.1, and hence is a Class 0 object.
It also has an estimated age of a few 10$^4$ years (AWB93).
We can see from Figure 3 that our model predicts that VLA1623 should have
accumulated less than 50\% of its mass at this at this age,
consistent with its Class 0 classification.
L1527 has a ratio of L$_{submm}$/L$_{bol}$ of 0.007, making it a borderline
Class 0/I object. It has an estimated age of $\sim$10$^5$ years (Ohashi et al.
1997). From Figure 3 we can see that at this age, our model would predict
that L1527 should have accumulated about 50\% of its mass, consistent
with its borderline Class 0/I categorization.

Thus we see that the model can reproduce easily the density profile and
dynamics of the pre-stellar core L1544, as well as reproducing the
statistical lifetimes of contracting
pre-stellar cores and Class 0 \& I protostars, and
the rapid growth of a protostar during the Class 0 phase. We
note once again that the model illustrated in Figures 1--3 is merely one
example of the generic class of Plummer-like models that we are presenting
here. Varying the central density $\rho\fl$
and the radius $R\fl$ produces a
family of solutions, and different
solutions may be appropriate for different star-forming regions.

\section{Model tests and comparison with other models}

The model we have developed here is simple, in the sense that it has 
effectively only 2 free parameters ($\rho\fl$ and $R\fl$). It represents 
well the observed density profiles of pre-stellar cores and Class 0
protostars. 
It reproduces the velocity field deduced for the pre-stellar core L1544,
and it matches the accretion rates and lifetimes inferred statistically for 
Class 0 and Class I protostars. It may therefore provide a useful framework 
for interpreting observations of pre-stellar cores and protostars, and 
for calculations of radiation transport and time-dependent chemistry.

Discriminating tests of the model will require further 
observations, such as larger surveys to improve the statistics 
upon which the time-scales and accretion rates are based, 
and more sensitive probes of the density and velocity profiles which must be 
reproduced. If such subsequent surveys continue to demonstrate what
the surveys thus far have indicated -- namely that the relative life-times of
Class~0 \& I protostars are as we have adopted here, and that the accretion 
rate decreases with time -- then this model will continue to prove useful.
If, however, more sensitive observations of pre-stellar cores show them to
have density and velocity profiles which (after modelling their collapse)
cannot reproduce the inferred protostellar life-times, then this would 
refute the model.

Our model is somewhat similar to 
that developed by Henriksen, Andr\'e 
\& Bontemps (1997), except that we do not take the density to be 
exactly uniform in the central region, and so we avoid an infinite 
accretion rate during the Class 0 phase. This has three advantages: 
first, it reduces the number of free parameters (our $\eta$ replaces
their $D_1$, $D_2$ and $r_{\mbox{\tiny b}}/r_{\mbox{\tiny N}}$); 
second, it allows us to fit the observationally inferred density 
and velocity profiles of pre-stellar cores; and third,
it enables us to use estimates of the duration of the Class 
0 phase to constrain the central density at the onset of collapse.

We note that the models of cores in thermal and mechanical 
equilibrium derived by Falgarone \& Puget (1985) and by Chi\`eze \& Pineau 
des For\^ets (1987) have structures very similar to what we 
are proposing, namely a flat density profile in the centre, and a 
steep fall-off further out with $\, - \, d \ell n [\rho] / d \ell n [r] 
\, \simeq \, 4 \,$. Similar density profiles are also predicted by 
ambipolar diffusion models (e.g. Ciolek \& Mouschovias 1994; Safier, 
McKee \& Stahler 1997), but these models have difficulty reproducing 
both the relatively short contracting
pre-stellar lifetimes and the relatively large inflow velocities.

Finally, we emphasize that, unless contraction is driven by external 
pressure (e.g. Myers \& Lazarian 1998), the freefall collapse of 
an initially flat-topped density configuration is almost certainly the 
least extreme way of delivering the rapid accretion rates inferred for 
Class 0 protostars. Any refinements to the model,
such as the inclusion of thermal pressure, 
turbulence, a magnetic field, or rotation,
are likely to slow the collapse down, in which case the pre-stellar 
core would have to start collapsing from a denser initial state.

\section{Conclusions}

In this paper we have proposed a new analytic model for the initial 
conditions of protostellar
collapse in moderately isolated regions of
star formation. The model is non-magnetic, and
is based on a Plummer-like 
radial density profile as its initial condition. 
The model fits: 

\begin{enumerate}

\item
the observed density profiles of pre-stellar cores;

\item
the observed density profiles of Class 0 protostars; 

\item
recent observations of roughly constant contraction velocities 
over a wide range of radii in the L1544 pre-stellar core;

\item
the lifetimes of contracting pre-stellar cores;

\item
the relative lifetimes
derived for Class 0 and Class I protostars;

\item
the relative inferred accretion 
rates of Class 0 and Class I protostars. 

\end{enumerate}

Moreover, the model is very simple, having in effect
only 2 free parameters, $\rho\fl$ and $R\fl$ ($\eta$ is so tightly
constrained by observations that it should not be considered `free').
Therefore it should provide a useful 
framework for interpreting observations of pre-stellar cores and protostars,
and for calculations of radiation transport and time-dependent chemistry.
We shall consider models involving external compression, rotation and magnetic 
fields in subsequent papers. We also plan to add radiative transfer
to our model to allow us to model line profiles directly. We would be happy
to supply copies of the model code upon request.\footnote{For a copy of the 
model code, email: ant@astro.cf.ac.uk}

\clearpage

\figcaption[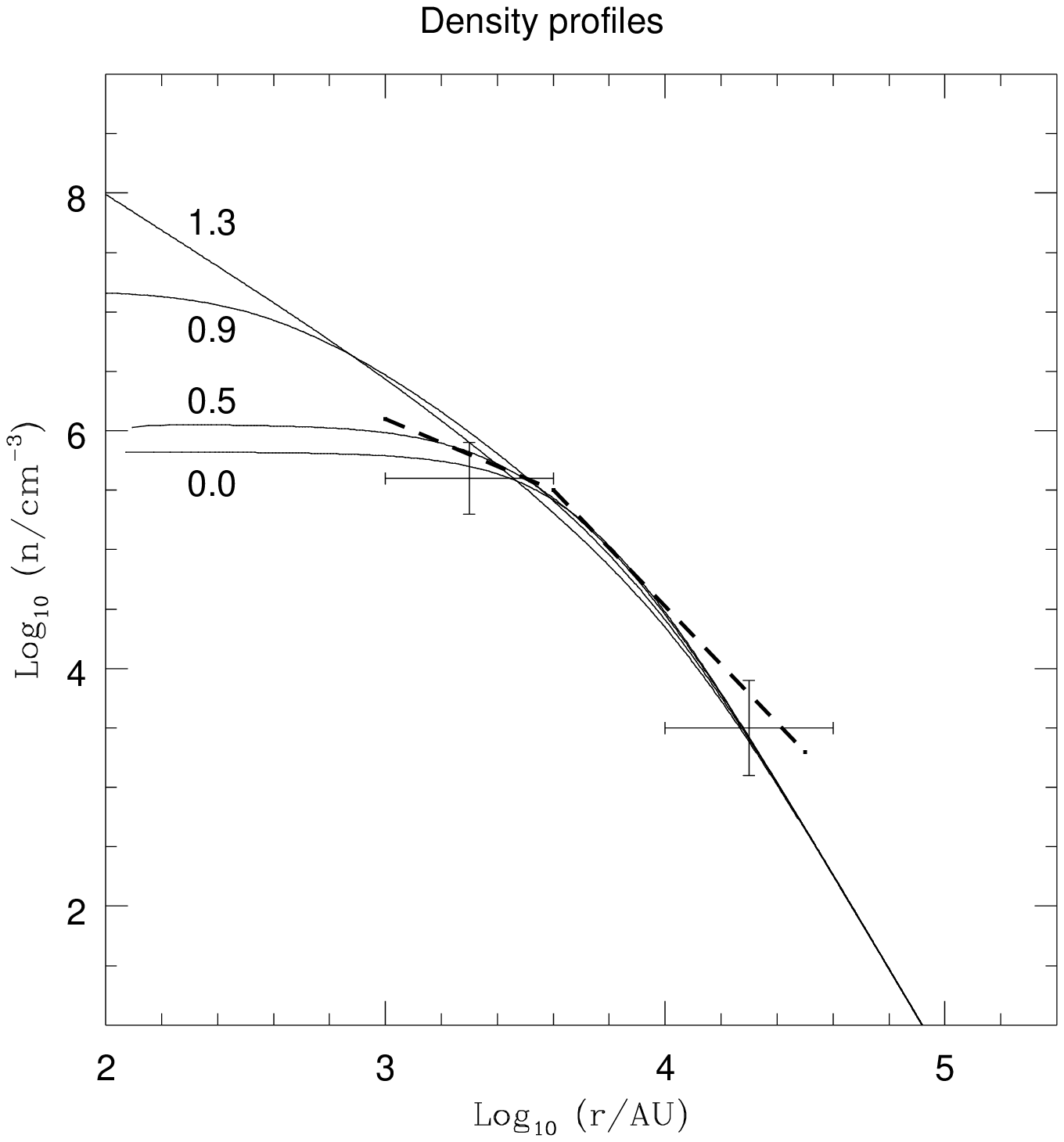]{Volume-density profiles. Each curve is marked with the 
value of $t/t\bg$, where the central point-mass forms at $t=t\bg$, so 
$t/t\bg < 1$ corresponds to contracting pre-stellar cores, and
$t/t\bg > 1$ corresponds to young Class 0 protostars.
The dashed line shows the density profile of L1544 
derived from IRAM 1.3~mm continuum mapping by Ward-Thompson et al. (1999). 
The crosses are the volume-density and radius 
values inferred by Tafalla et al. (1998) 
and Williams et al. (1999) for L1544. 
Hence all of these observations of L1544 
are consistent with $t/t\bg \sim$ 0.5. \label{fig1}}

\figcaption[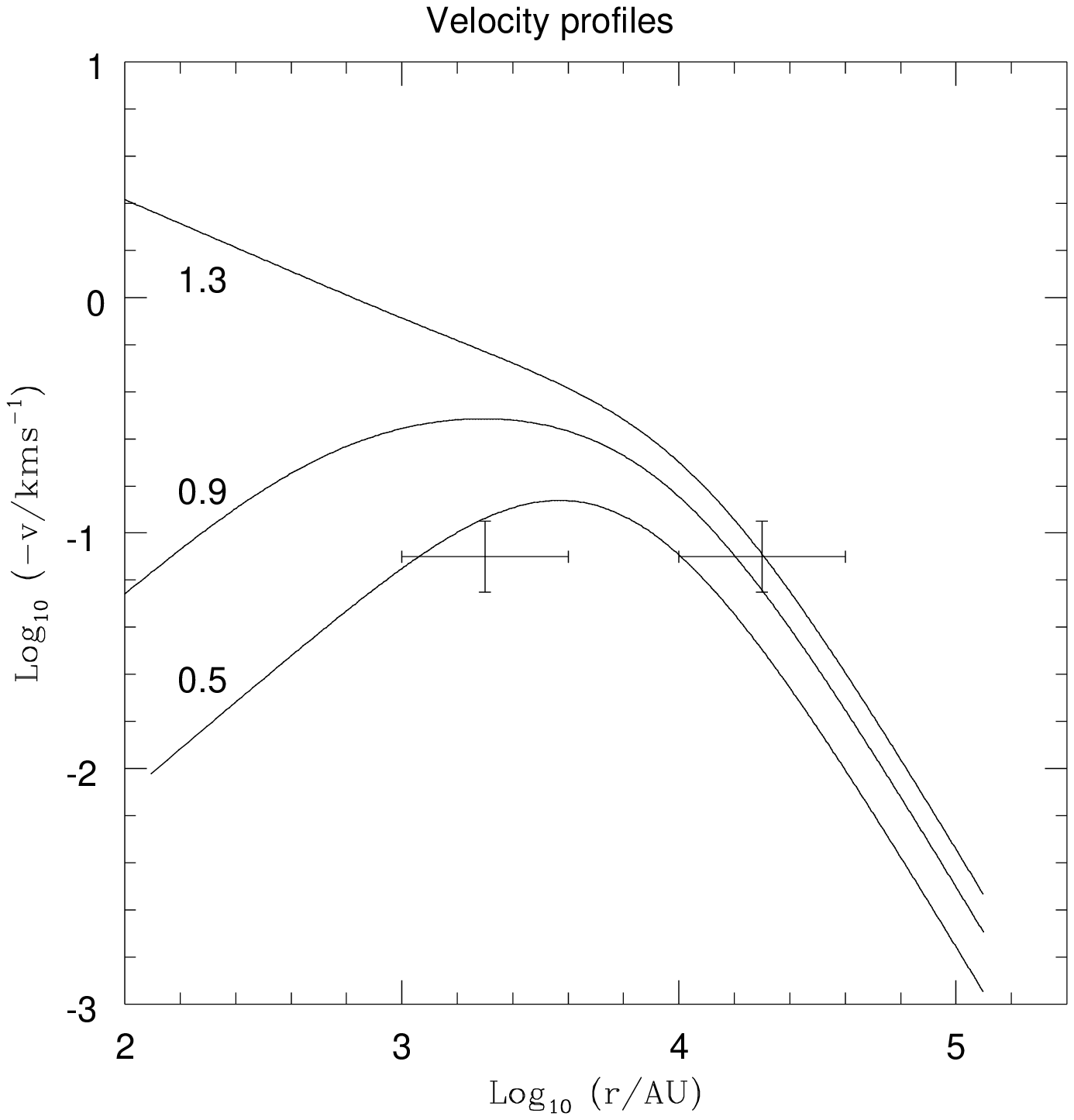]{Inward radial velocity profiles. Each curve is marked 
with the value of $t/t\bg$, as in Figure 1. The crosses represent 
the infall velocities inferred by Tafalla et al. (1998) 
and Williams et al. (1999) for L1544. Once again a value
of $t/t\bg \sim$ 0.5 is consistent with the observations. \label{fig2}}

\figcaption[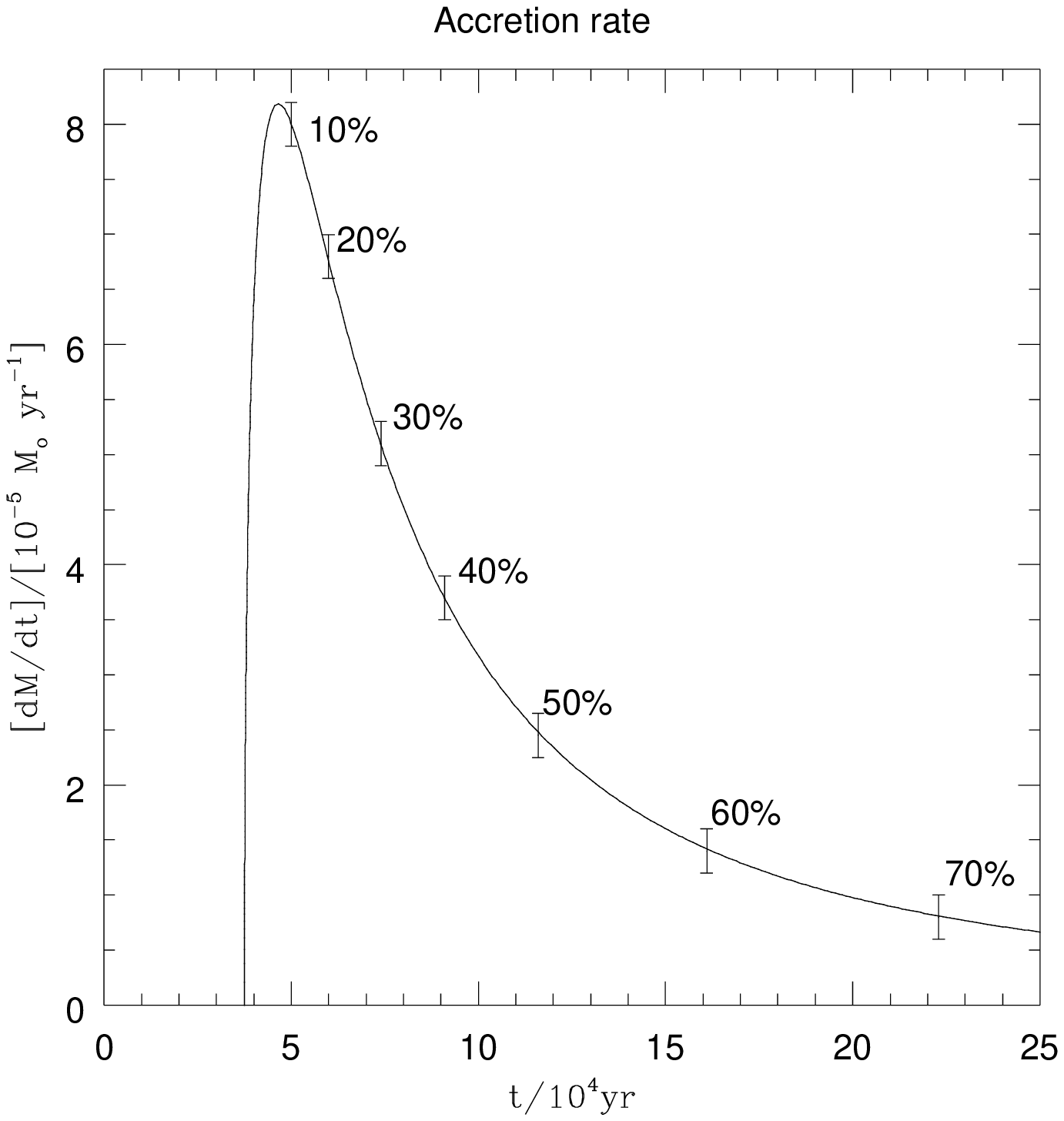]{The 
accretion rate, in units of $10^{-5} M_\odot$ yr$^{-1}$, 
plotted against time, in units of $10^4$ yrs. 
The borderline between pre-stellar cores and Class 0 protostars occurs
when the accretion rate first becomes non-negligible.
The tick marks indicate the 
fraction of the total mass which has reached the central protostar.
The Class 0/I borderline occurs when this fraction is 50\%.
\label{fig3}}

\begin{figure}
\setlength{\unitlength}{1mm}
\begin{picture}(80,120)
\includegraphics{f1.eps}
\end{picture}
\end{figure}

\begin{figure}
\setlength{\unitlength}{1mm}
\begin{picture}(80,120)
\includegraphics{f2.eps}
\end{picture}
\end{figure}

\begin{figure}
\setlength{\unitlength}{1mm}
\begin{picture}(80,120)
\includegraphics{f3.eps}
\end{picture}
\end{figure}

\end{document}